\documentstyle[12pt]{article}
\newcommand{\beq}{\begin{equation}}
\newcommand{\eeq}{\end{equation}}
\newcommand{\beqa}{\begin{eqnarray}}
\newcommand{\eeqa}{\end{eqnarray}}

\newcommand{\bma}{\left( \begin{array}}
\newcommand{\ema}{\end{array} \right)}
\newcommand{\bfig}{\begin{figure}}
\newcommand{\efig}{\end{figure}}
\newcommand{\bc}{\begin{center}}
\newcommand{\ec}{\end{center}}
\newcommand{\pslash}{\kern 0.2 em p\kern -0.45em /}
\newcommand{\sla}[1]{\kern 0.2 em #1\kern -0.45em /}

\begin{document}
\setcounter{page}{0}
\thispagestyle{empty}
\hspace*{12.0cm}                        WU-B 94-17\\
\hspace*{12.0 cm}                     August 1994\\
\bc
{\Large\bf ELECTROMAGNETIC FORM FACTORS AT LARGE MOMENTUM TRANSFER }\\
\vspace*{1.0 cm}
\ec
\bc
{\large P. Kroll}
\footnote{Supported in part by BMFT, FRG under contract 06 WU 737}
\footnote{E-mail: kroll@wpts0.physik.uni-wuppertal.de}
\footnote{Invited talk presented at
           the Intern.~Conf.~on Physics with GeV-Particle Beams,
J\"ulich (1994)}\\
\vspace*{0.5 cm}
Fachbereich Physik, Universit\"{a}t Wuppertal, Gau\ss strasse 20,\\
D-42097 Wuppertal 1, Germany\\[0.3 cm]
\ec
\vspace*{4.0 cm}
\bc
                    {\bf Abstract}
\ec
Recent improvements of the hard scattering picture for the large $p_{\perp}$
behaviour of electromagnetic form factors, namely the inclusion of both
Sudakov corrections and intrinsic transverse momentum dependence of the
hadronic wave function, are reviewed. On account of these improvements
the perturbative contributions to the pion's and the nucleon's form factor
can be calculated in a theoretically self-consistent way for momentum
transfers as low as about $2$ and $3\,{\rm GeV}$, respectively.
This is achieved at the expense of a substantial
suppression of the perturbative contribution in the few GeV region.
Eventual higher twist contributions are discussed in some detail.\\
\newpage
\section{The hard scattering picture}
\setcounter{equation}{0}
\vspace*{-0.5cm}
There is general agreement that perturbative QCD in the framework of the
hard-scattering picture (HSP) is the correct description of form factors
at asymptotically large momentum transfer (see \cite{lep:80} and references
therein). In the HSP a form factor is expressed by a convolution of
distribution
amplitudes (DA) with hard scattering amplitudes calculated in collinear
approximation within perturbative QCD. The universal, process independent
DAs, which represent hadronic wave functions integrated over
transverse momenta, are controlled by long distance physics
in contrast to the hard scattering amplitudes which are governed by
short distance physics. The DAs cannot be calculated by perturbative
means, we have to rely on models. In principle lattice gauge theory
offers a possibility to calculate the DAs but with the present-day
computers a sufficient accuracy can not be achieved, only a few moments
of the pion and the proton DA have been obtained \cite{mar:88}. It is of
utmost phenomenological interest whether or not the asymptotic
perturbative result can already be applied at experimentally
accessible momentum transfers. The major topic of this talk is to answer
that question. In order to keep
the technical effort simple I am going to discuss the electromagnetic
form factor of the pion mainly. The generalization to the phenomenological
more important case of the nucleon form factor is straightforward.\\
Now let us consider the electromagnetic form factor of the pion.
To lowest order pertubative QCD the hard scattering amplitude $T_H$ is
to be calculated from the two one-gluon exchange diagrams.
Working out the diagrams one finds
\beq
\label{eq:hs-amplitude}
T_H (x_1,y_1,Q,\vec k_\perp,\vec l_\perp) =
\frac{16 \pi \, \alpha_s(\mu) \, C_F}
{x_1 y_1 Q^2+(\vec k_\perp + \vec l_\perp)^2},
\eeq
where $Q (\geq 0)$ is the momentum transfer from the initial to the final
state pion. $x_1$ ($y_1$) is the longitudinal momentum fraction carried by
the quark and $\vec{k}_{\perp}$ ($\vec{l}_{\perp}$) its transverse momentum
with respect to the initial (final) state pion. The momentum of the
antiquark is characterized by $x_2=1-x_1$ ($y_2=1-y_1$) and
$-\vec{k}_{\perp}$ ($-\vec{l}_{\perp}$). $C_F$ ($=4/3$) is the colour
factor and $\alpha_s$ is the usual strong coupling constant to be evaluated
at a renormalization scale $\mu$. The expression (\ref{eq:hs-amplitude})
is an approximation in so far as only the most important $\vec{k}_{\perp}$-
and $\vec{l}_{\perp}$-dependences have been kept. Denoting the wave
function of the pion's valence Fock state by $\Psi_0$, the form factor
is given by
\beq
\label{eq:pre-hs-Fpi}
F_\pi(Q^2)=
\int \frac{dx_1 \, d^{\;\!2} k_\perp}{16 \pi^3}
\int \frac{dy_1 \, d^{\;\!2} l_\perp}{16 \pi^3} \,
\,\Psi_0^\ast (y_1,\vec l_\perp)\, T_H (x_1,y_1,Q,\vec k_\perp,\vec l_\perp)
\, \Psi_0 (x_1,\vec k_\perp).
\eeq
Strictly speaking $\Psi_0$ represents only the soft part of the pion
wave function, i.e. the full wave function with the perturbative tail
removed from it \cite{lep:80}. Contributions from higher Fock states are
neglected in (\ref{eq:pre-hs-Fpi}) since, at large momentum transfer, they
are suppressed by powers of $\alpha_s/Q^2$.\\
At large $Q$ one may neglect the $k_\perp$- and $l_\perp$-dependence
in the gluon propagator as well; $T_H$ can then be pulled out of
the transverse momentum integrals, and these integrations apply only to the
wave functions. Defining the DA by
\beq
\label{eq:DAdef}
\frac{f_\pi}{2 \sqrt{6}} \,\, \phi (x_1,\mu_F) =
\int \frac{d^{\;\!2} k_\perp}{16 \pi^3}\,  \Psi_0 (x_1,\vec k_\perp),
\qquad\qquad
\int_0^1 dx_1 \, \phi(x_1,\mu_F) = 1,
\eeq
one arrives at the celebrated hard-scattering formula for the pion's
form factor
\beq
\label{eq:hs-Fpi}
{F_\pi}^{HSP}(Q^2)=\frac{{f_\pi}^2}{24} \int dx_1 \, dy_1
\, \phi^\ast(y_1,\mu_F) \, T_H(x_1,y_1,Q,\mu) \, \phi(x_1,\mu_F),
\eeq
which is valid for $Q \to \infty$.
$f_\pi \,(=133\, {\rm MeV})$ is the usual $\pi$ decay constant.
$\mu_F$ is the scale at which the physics is factorized into its
soft and hard components. The DA mildly depends on the factorization scale
(QCD evolution).\\
An appropriate choice of the
renormalization scale is $\mu =\sqrt{x_1 y_1} Q$. This avoids large logs
from higher order pertubation theory at the expense, however, of a
singular behaviour of $\alpha_s$ in the end-point regions where one of
the fractions, $x_i$ or $y_i$ ($i=1,2$), tends to zero.
It is argued that radiative corrections (Sudakov factors) will
suppress that singularity and, therefore, in practical applications
of the HSP one may handle that difficulty by freezing $\alpha_s$
at a certain value, which is typically chosen in the range of 0.5 to 0.7.
That crude recipe is unsatisfactory although the Sudakov argument itself is
correct as will be discussed in the next section. Another often used and
very convenient recipe is to pull out of the integral in (\ref{eq:hs-Fpi})
the coupling $\alpha_s$ and to take a suitable fraction of the
momentum transfer as its argument. Both the recipes have the unwelcome
consequence of introducing an external free parameter in the calculation.
The second recipe for the treatment of $\alpha_s$ allows one to cast the
HSP prediction for the pion form factor into the simple form
\beq
\label{eq:fmom}
Q^2 F_{\pi}^{HSP} = \frac{2\pi}{3} f^2_{\pi} C_F \alpha_s(\mu)
                         \langle x_1^{-1}\rangle ^2
\eeq
which nicely brings to light that the moment $\langle x_1^{-1}\rangle$
($=\int dx_1 \Phi (x_1) x_1^{-1}$) is the only information on the DA actually
required, a fact which approximately remains true if $\alpha_s$
is kept in the integral. One also reads off from (\ref{eq:fmom}) that
the HSP predicts a behaviour of the form factor as
$\sim 1/Q^2 ({\rm modulo}\ln Q)$ in agreement with dimensional counting.\\
In the formal limit $Q^2 \to \infty$ the HSP even determines the moment
$\langle x_1^{-1}\rangle$
and hence the value of the form factor (except
of the uncertainty in the treatment of $\alpha_s$). It can be shown
\cite{lep:80} that any DA evolves into the asymptotic form
$\Phi_\pi^{AS} = 6 x_1 x_2$ which entails $\langle x_1^{-1}\rangle = 3$ and,
using $\alpha_s(\mu)=0.33$, $Q^2 F_{\pi}^{HSP}=0.15$. This prediction is too
small by about a factor of 2 as compared with the admittedly poor
data \cite{Beb:76}. In order to obtain
better results a larger value of the moment $\langle x_1^{-1}\rangle$
is required. Obviously, this can be achieved with DAs being strongly
concentrated in the end-point regions. Such DAs, first
proposed by Chernyak and Zhitnitsky (CZ) \cite{CZ:82}, find a certain
justification in QCD sum rules by means of which a few moments of
the DAs have been calculated. CZ suggested the DA
$\Phi_\pi^{CZ} = 30 x_1 x_2 (x_1-x_2)^2$
yielding $\langle x_1^{-1}\rangle = 5$ and hence a value for the form factor
in accord with the data. The CZ moments are subject to considerable
controversy: Other QCD sum rule studies provide different values for
the moments \cite{rad:91}. Also the results obtained from lattice
gauge theory do not well agree with the CZ moments \cite{mar:88}. Finally,
the data on the $\pi - \gamma$ transition form factor, which is also determined
by the moment $\langle x_1^{-1}\rangle$, favour the asymptotic DA.  \\
The magnetic form factor of the nucleon can also be analysed within the HSP.
The calculations reveal a dramatic dependence of the results on the
utilized DA. Asymptotically, i.~e.~with the DA $\sim 120 x_1 x_2 x_3$, one
obtains
\beq
\label{nucasy}
 G_M^p = 0, \qquad\qquad\qquad Q^4\;G_M^n = \left( \frac{\alpha_s(Q))}
        {\alpha_s(\mu_0)} \right) ^{4/3\beta_0}  f_N^2(\mu_0) \frac{100}{3}
        (4 \pi \alpha_s(Q))^2
\eeq
where $\beta_0 = 11-2/3 n_f = 9$ and $\mu_0 \simeq 1\, \rm{GeV}$.
$f_N(\mu_0) $ ($= (5.0 \pm 0.3)\times 10 ^{-3}\, \rm{GeV}^2$) represents
the value of the nucleon wave function at the origin in configuration
space. The factor in front of $f_N$ takes into account the evolution of $f_N$.
These results do not bear any resemblance to the experimental data
($Q^4 G_M^p \sim 1\, \rm{GeV}^4$ for $Q^2$ between $10$ and $30\, \rm{GeV}^2$,
see \cite{sil:93}). Again, as in the pion case,
strongly end-point concentrated DAs yield good results
provided an appropriate value is chosen for $\alpha_s$.
A set of such end-point concentrated DAs which all respect the QCD sum rules
constraints, has been determined by Bergmann and Stefanis (BS)\cite{BS:93}.\\
The Pauli form factor $F_2$, and hence the electric form factor $G_E$,
cannot be calculated within the HSP since it requires helicity-flip
transitions which are not possible for (almost) massless quarks in the
collinear approximation. The Pauli form factor is dominated by sizeable
higher twist contributions in the few GeV region as we know from
experiment \cite{bos}. The higher twist nature of the Pauli form factor
is clearly visible in Fig.~1: its large $Q$ behaviour is compatible
with $1/Q^6$.
\vspace*{-0.7cm}
\section{The Botts-Li-Sterman approach}
\setcounter{equation}{0}
\vspace*{-0.5cm}
The applicability of the HSP at experimentally accessible momentum
transfers, typically a few GeV, was questioned \cite{rad:91,Isg:89}. It
\bfig[p]
\vspace*{-0.5cm}
\caption[abcd]{\small
The Pauli form factor of the proton scaled by $Q^6$. Data are taken
from {\rm\cite{bos}}. The solid line represents the results obtained
from the diquark model {\rm\cite{jkss:93}}.}
\efig
was asserted that in the few GeV region the hard-scattering picture
accumulates large contributions from the end-point regions where the parton
virtualities are small. This renders the perturbative calculation inconsistent,
in particular for the end-point concentrated DAs.
The use of the collinear approximation, i.~e.~the neglect of the
transverse momentum dependence of the hard scattering amplitude,
see for instance (\ref{eq:hs-amplitude}), is also unjustified in
the end-point regions. Obviously, the collinear approximation entails large
errors in the final results for the end-point concentrated DAs whereas
for the asymptotic DA and similar forms it turns out to be reasonable. \\
The statements made by the authors of \cite{rad:91,Isg:89} were
challenged by Sterman and collaborators \cite{bot:89,LiS:92,Li:92}.
These authors suggest to retain the transverse momentum dependence
of the hard scattering amplitude and to take into account Sudakov
corrections. In order to include the Sudakov corrections it is advantageous to
reexpress (\ref{eq:pre-hs-Fpi}) in terms of the Fourier conjugated
variable $\vec b$ in the transverse configuration space
\beq
\label{eq:ft-Fpi}
{F_\pi}^{pert}(Q^2) =
\int\! \frac{dx_1 \, dy_1}{(4 \pi)^2}
\int\!d^{\;\!2}b
\,\hat{\Psi}_0^\ast (y_1,\vec b)\,\hat{T}_H (x_1,y_1,b,Q,t)
\,\hat{\Psi}_0 (x_1,-\vec b) \,\exp\left[-S\right]
\eeq
where the Fourier transform of a function $f=f(\vec k_\perp)$ is
denoted by $\hat{f}=\hat{f}(\vec b)$. As the renormalization scale
Sterman et al. choose the largest mass scale appearing in $\hat{T}_H$, the
Fourier transform of the lowest order hard scattering amplitude
(\ref{eq:hs-amplitude}):
\beq
\label{eq:als-arg}
t= {\rm Max}(\sqrt{x_1 y_1}\,Q,1/b).
\eeq
The factor $\exp \left[-S\right]$ in
(\ref{eq:ft-Fpi}), termed the Sudakov factor, incorporates the effects
of gluonic radiative corrections and therefore represents parts of higher order
perturbative corrections to $T_H$. Botts and Sterman \cite{bot:89} have
calculated the Sudakov factor using resummation techniques and having
recourse to the renormalization group. For the pion case they find a
Sudakov exponent of the form
\beq
\label{sud:1}
S(x_1, y_1, b, Q, t) = \sum_{i=1}^{2} [ s(x_i, b, Q) + s(y_i, b, Q)]
                      -\frac{8}{3 \beta_0} \ln \frac{\ln(t/\Lambda_{QCD})}
                       {\ln(1/b\Lambda_{QCD})}
\eeq
The lengthy expression for the Sudakov function $s=s(\xi_i, b, Q)$,
which includes all leading and next-to-leading logarithms, is given explicitly
in \cite{LiS:92}. The most important term in it is the double logarithm
\beq
\label{eq:double-log}
\frac{8}{3\beta_0} \ln \frac{\xi_i Q}{\sqrt{2} \Lambda_{QCD}}
\ln \frac{\ln (\xi_i Q/\sqrt{2} \Lambda_{QCD})}{\ln (1/b \Lambda_{QCD})},
\eeq
where $\xi_i$ is one of the fractions, $x_i$ or $y_i$. The quark-antiquark
transverse separation acts as an infrared cut-off. The underlying physical
idea is the following: Because of the colour neutrality of a hadron,
its quark distribution cannot be resolved by a gluon with a wave length
much larger than the $q - \bar{q}$ separation. Consequently radiation
is damped. The infrared cutoff marks the interface between the
non-perturbatively soft momenta, which are implicitly accounted for in the
hadronic wave function, and the contributions from semi-hard gluons,
incorporated in a perturbative way in the Sudakov factor. Whenever
$1/b$ is large relative to the hard (gluon) scale $\xi_i Q$, the gluonic
corrections are to be considered as hard gluon corrections to $\hat{T}_H$ and
hence are not contained in the Sudakov factor but absorbed in $\hat{T}_H$. For
that reason the Sudakov function $s(\xi_i, b, Q)$ is set equal to zero
whenever $\xi_i \leq \sqrt{2}/b Q$.\\
The crucial advantage of the modified HSP (\ref{eq:ft-Fpi}) is that the
renormalization scale can now be chosen to be $\mu = \sqrt{x_1 y_1} Q$ and
so large logs from higher order perturbation theory be avoided. The singularity
of the ``bare'' $\alpha_s$ is suppressed by the Sudakov factor
inherently; there is no need for external regulators! This can be observed
from Fig.~2 where the exponential of the Sudakov function
$\exp{[-s(\xi_i, b, Q)]}$ is displayed. For small $b$ there is no suppression.
As $b$ increases $\exp{[-s]}$ decreases and drops to zero faster than any power
of $\ln{(1/b\Lambda_{QCD})}$ for $b\Lambda_{QCD} \to 1$ except one is
in the dangerous region, $\xi_i \leq \sqrt{2}/b Q$, where $s(x_i, b, Q)$
is set to zero. However, in this case $\exp{[-s(1-\xi_i, b, Q)]}$ provides
the required suppression. Consequently, the Sudakov factor (\ref{sud:1}) drops
to zero faster than any power of $\ln{(1/b\Lambda_{QCD})}$ for
$b\Lambda_{QCD} \to 1$
irrespective of the value of $\xi_i$. This behaviour of the Sudakov factor
guarantees the cancellation of the $\alpha_s$ singularity (owing to the
limit $t \to \Lambda_{QCD}$).\\
For $b\Lambda_{QCD}$ larger than 1, considered as the true soft region,
the Sudakov factor is set to zero. Note that for $Q \to \infty$ the Sudakov
factor damps any contribution except those from configurations with
small quark-antiquark separations. In other words, the hard-scattering
contribution (\ref{eq:hs-Fpi}) dominates the form factor asymptotically.\\
\bfig[p]
\caption[bcde]{\small
The exponential of the Sudakov function $s(\xi_l,b,Q)$ vs.~$\xi_l$ and
$\tilde {b}_l\Lambda_{QCD}$ for $Q=30\,\Lambda_{QCD}$. In the pion case
$\tilde{b}_l$ equals $b$. In the hatched area the Sudakov function is
set equal to zero.}
\efig
\vspace*{-1.0cm}
\section{The intrinsic $k_\perp$-dependence of the wave function}
\setcounter{equation}{0}
\vspace*{-0.5cm}
The approach proposed in \cite{bot:89,LiS:92,Li:92}
certainly constitutes an enormous progress in our understanding of
exclusive reactions at large momentum transfer.
In any practical application of that approach one has however to allow
for an intrinsic transverse momentum dependence of the hadronic wave
function \cite{jak:93}, although, admittedly, this requires a new
phenomenological element in the calculation. Fortunately, in the case of
the pion the intrinsic transverse momentum of its valence Fock
state wave function is rather well constrained. In accordance with
(\ref{eq:DAdef}) the wave function can be written as
\beq
\label{eq:wvfct}
\Psi_0 (x_1,\vec k_\perp) = \frac{f_\pi}{2 \sqrt{6}}
\,\phi(x_1) \,\Sigma(x_1,\vec k_\perp),
\eeq
the function $\Sigma$ being normalized in such a way that
\beq
\label{eq:Sigmanorm}
\int \frac{d^{\;\!2} k_\perp}{16 \pi^3} \,\Sigma (x_1,\vec k_\perp) =1.
\eeq
The wave function (\ref{eq:wvfct}) is subject to the following
constraints: it is normalized to a number $P_{q\bar{q}}\leq 1$, the
probability of the valence quark Fock state; the value of the
configuration space wave function at the origin is determined by the
$\pi$ decay constant; the process $\pi ^0 \to \gamma\gamma$ provides a
third relation. Finally, the charge radius of the pion provides a lower
limit on the root mean square (r.m.s.) transverse momentum; actually it
should be larger than $300 \,{\rm MeV}$.
The $k_\perp$-dependence of the wave function is parameterized as
a simple Gaussian
\beq
\label{eq:Sigma}
\Sigma(x_1,\vec k_\perp)=16 \pi^2 \beta^2 \,g(x_1)
\,\exp \left(-g(x_1) \beta^2 k_\perp^2 \right),
\eeq
$g(x_1)$ being either  $1$ or $1/x_1x_2$. The latter case goes along
with a factor $\exp (-\beta^2 m_q^2/x_1 x_2 ) $ in the DA
where $m_q$ is a constituent quark mass ($330 \,{\rm MeV}$). The Gaussian
(\ref{eq:Sigma}) is consistent with the required large-$k_\perp$ behaviour
of a soft wave function. Several wave functions have been employed in
\cite{jak:93}. Here, in this talk, only the results for the two
extreme cases utilized in \cite{jak:93} are quoted. That is, on the one hand,
the CZ wave function $\sim \Phi_\pi^{CZ}$ , $g=1$ \cite{CZ:82}
which is the example most concentrated in the end-point region and,
on the other hand, the modified asymptotic (MAS) wave function
$\sim \Phi_\pi^{AS}$, $g=1/x_1x_2$. The MAS wave function is the
example least concentrated in the end-point regions.
\bfig[p]
\caption[figname]{\small
(Left) The pion's form factor as a function of $Q^2$ evaluated
with the CZ wave function and $\Lambda_{QCD}=200\,{\rm MeV}$. The dash-dotted
line is obtained from (\ref{eq:hs-Fpi}) with $\alpha_s$ frozen
at 0.5 and the dashed line from {\rm(\ref{eq:ft-Fpi})} ignoring the intrinsic
$k_\perp$-dependence \cite{LiS:92}. The solid line represents the complete
result obtained from {\rm(\ref{eq:ft-Fpi})}
($\langle k_\perp^2 \rangle^{1/2}_{\Psi} =350\,{\rm MeV}$).
Data are taken from {\rm\cite{Beb:76}} ($\circ$ 1976, $\bullet$ 1978).\\
(Right) As left figure but using the MAS wave function. Note the modified
scale of the abscissa.}
\efig
The parameter $\beta$ in the wave function is fixed by requiring specific
values for the r.m.s. transverse momentum. For a value of $350\,{\rm MeV}$
all the constraints on the pion wave functions are well respected
\cite{jak:93}.\\
Li and Sterman \cite{LiS:92} assume that the dominant $b$-dependence
of the integrand in (\ref{eq:ft-Fpi}) arises from the Sudakov
factor and that the Gaussian in the Fourier transformed $\Sigma$ can
consequently be replaced by 1. Numerical evaluations of the pion's form factor
through (\ref{eq:ft-Fpi}), using the various wave functions mentioned
above, reveal that this assumption leads to an overestimate of the perturbative
contribution (see Fig.~3). For momentum transfers of the order of a few GeV the
wave function damps the integrand in (\ref{eq:ft-Fpi}) more than
the Sudakov factor which takes over only for very large
values of $Q$. The suppression caused by the intrinsic
transverse momentum dependence of the wave function,
is particular strong for the end-point concentrated
wave functions. These observations confirm the statements made
at the beginning of Sect.~2: the so-called success of the end-point
concentrated DAs is only fictitious; for finite values of $Q$ the HSP formula
(\ref{eq:hs-Fpi}) does not represent a reasonable approximation
to (\ref{eq:pre-hs-Fpi}) for such DAs.\\
The numerical studies reveal that the modified hard scattering approach is
self-consistent for $Q\geq 2 {\rm GeV}$ in the sense that less than, say,
$50\%$ of the result is generated by soft gluon exchange $(\alpha_s > 0.7)$.
The exact value of $Q$ at which self-consistency sets in
depends on the wave function. It is larger for the end-point concentrated wave
functions than for the MAS DA. However, the perturbative
contribution (\ref{eq:ft-Fpi}), although self-consistent, is
presumably too small as compared with the (poor) data. \\
The Gaussian $k_{\perp}$ dependence of the wave function is a plausible
but special case and one may suspect that more complicated functions
lead to a better agreement with the data. Recently, in QCD sum rule analyses
similar to those determining the moments of the DAs $\langle x_1^n\rangle$,
the lowest $k_{\perp}$-moments, $\langle k_{\perp}^n\rangle$ have
been estimated \cite{zhi:94,lee:94}. The ratio
$\langle k_{\perp}^4\rangle /\langle k_{\perp}^2\rangle ^2$
is found to have a value in the range $5-8$. For comparison the Gaussian
(\ref{eq:Sigma}) provides a value of about 2.
The large value of the ratio implies a $k_{\perp}$ distribution being not
strongly concentrated around the point $k_{\perp} =0$. Modelling a wave
function which respects that new constraint, e.~g.~a Gaussian plus
a $\delta$-function, does however not lead to a substiantally larger
perturbative contribution to the pion form factor.
\vspace*{-0.7cm}
\section{The magnetic form factor of the nucleon}
\setcounter{equation}{0}
\vspace*{-0.5cm}
The nucleon's magnetic form factor $G_M$ can be analysed within the
modified HSP along the same lines as the pion's form factor.
The basic formula for the perturbative contribution is an expression
similar to (\ref{eq:ft-Fpi}). Since the nucleon's valence Fock state
consists of three quarks, one has two independent $x$ (and $y$) variables
and two independent transverse interquark separations. The wave function
is parameterized analogously to (\ref{eq:wvfct}) with the DA being a
linear combination of Appell polynomials $\tilde{\Phi}^n$ which are the
eigenfunctions of the evolution equation \cite{lep:80}
\beq
\label{appell}
\Phi(x_1,x_2,x_3)=120 x_1 x_2 x_3 \sum_n B_n \left(\frac{\alpha_s(\mu_F)}
          {\alpha_s(\mu_0)}\right)^{\tilde{\gamma}_n/\beta_0}
          \tilde{\Phi}^n(x_1,x_2,x_3).
\eeq
The exponents $\tilde{\gamma}_n$, driving the
evolution behaviour of the DA, are positive fractional numbers increasing
with $n$. For the BS set of DAs \cite{BS:93} the sum in (\ref{appell}) is
truncated at $n=5$ (corresponding to the polynomial order 2) and the
coefficients $B_n$ are determined in the context of QCD sum rules.\\
Now the Sudakov exponent $S$ consists of a sum of six Sudakov functions
$s(\xi_l,\tilde{b}_l,Q)$ ($\xi_l=x_l$ or $y_l$, $l=1,2,3$) and of the
integral over the anomalous dimensions (see (\ref{sud:1})) which depends on the
mass scales associated with the hard gluons. The choice of the infrared
cut-off parameters $\tilde{b}_l$ is not as simple as in the pion case.
These parameters are naturally related to but not uniquely determined
by the mutual separations of the three quarks. Li, in his pioneering analysis
of the proton form factor within the modified HSP \cite{Li:92}, chooses
$\tilde{b}_l=b_l$ where $b_1$ ($b_2$) is the transverse separation of
quark 1 (2) and 3 and $b_3$ that of quark 1 and 2. However, Li's
analysis is seriously flawed: the cancellation of the $\alpha_s$ singularities
for $\tilde{b}_l \Lambda_{QCD} \to 1$ ($x_l$ fixed) is incomplete for
his choice of the infrared cut-offs. The uncompensated singularites are
of the form
\beq
\label{sing}
\sim \ln ^{-\kappa}(1/\tilde{b}_l\Lambda_{QCD})
\eeq
The maximum degree of divergence $\kappa$ is $7/9$ if evolution is
ignored but larger than $123/81$ if it is taken into account (note, the
factorization scale $\mu_F$ is $\min\{1/\tilde{b}_l\}$).
The Sudakov factor does not necessarily vanish fast enough to guarantee
the cancellation of these singularities. Consider for example the string like
configurations: $x_1 \le \sqrt{2}/(\tilde{b}_1 Q)$ and
$\tilde{b}_1\Lambda_{QCD}\to 1$, the region where $s(x_1,\tilde{b}_1,Q)$ is
set to zero, and
$\tilde{b}_2\Lambda_{QCD}\simeq\tilde{b}_3\Lambda_{QCD}\simeq 1/2$. Then,
none of the other two Sudakov functions, $s(x_2,\tilde{b}_2,Q)$ and
$s(x_3,\tilde{b}_3,Q)$, tends to infinity and hence the Sudakov factor
does not provide any suppresssion.\\
In a recent paper another choice for the infrared cut-offs has been
suggested \cite{bol:94a}, namely a common cut-off
$\tilde{b}=\tilde{b}_1=\tilde{b}_2=\tilde{b}_3=\max \{b_l\}$. Obviously,
for this choice, termed the ``MAX'' prescription, at least one of the
Sudakov functions tends to infinity if $b_l\Lambda_{QCD} \to 1$
(at least one of the $x_l$ is larger than $\sqrt{2}/\tilde{b}Q$). The ``MAX''
prescription does not only lead to a regular integral but also to a
non-singular integrand which is a prerequisite for the self-consistency of
the modified HSP.\\
The numerical results for the proton's magnetic form factor are compared
to the data \cite{sil:93} in Fig.~4. The hatched band indicates the range
of predictions derived for the BS set of DAs. Each DA is multiplied by a
Gaussian of the type (\ref{eq:Sigma}) and the full wave function is normalized
to unity fixing the parameter $\beta$. The values for the r.~m.~s. transverse
momentum range from $270$ to $320\,{\rm MeV}$ for the various wave functions.
These values are to be considered as minimum values, they may be
larger (providing stronger suppression of the perturbative contribution).
The band of predictions does not overlap with the data, the predictions are
too low by about a factor of 2 to 3. Even if the intrinsic transverse
momentum is ignored there is no overlap.
\bfig[p]
\caption[bcde]{\small
The proton's magnetic form factor vs. $Q^2$. Data are taken from \cite{sil:93}
(filled (open) circles $G_M$ ($F_1$). Theoretical results are obtained with
the ``MAX'' prescription using the DAs given in \cite{BS:93}
($\Lambda_{QCD}=180\,{\rm MeV}$). The wave functions are normalized to unity. }
\efig
The perturbative contribution becomes self-consistent for $Q\geq 3\,{\rm GeV}$:
$50\%$ of the result is accumulated in regions where $\alpha _{s}^{2}\leq 0.5
$.
The onset of self-consistency mildly depends on the wave function used. \\
An analysis of the neutron's magnetic form factor \cite{bol:94a} yields results
similar in trend but in apparently better agreement with the data.
There is overlap between the band of predictions and the data at the largest
measured momentum transfers where, incidentally, the theoretical calculation
becomes self-consistent.\\
Finally, it should be noted that Hyer \cite{hy:93} has investigated the
magnetic form factor of the proton in the time-like region within the
modified HSP. Comparing with the data of the Fermilab E760 collaboration
\cite{arm:93} (see Fig.~5), also Hyer's result is too small.
\vspace*{-0.7cm}
\section{Soft contributions to the form factors}
\setcounter{equation}{0}
\vspace*{-0.5cm}
As discussed in sections 3 and 4 the perturbative contributions to
the pion's and the nucleon's form factor are too small. Hence
other contributions must play an important role
in the few GeV region. Obviously, for a perturbative calculation one may
suspect higher order contributions to be responsible for the discrepancy
between theory and experiment. In analogy to the Drell-Yan process
such contributions may be condensed in a K-factor multiplying the
lowest order result for the form factor\\
\beq
\label{kfac}
K = 1 + \frac{\alpha_s (\mu)}{\pi} B(Q, \mu) + {\cal O} (\alpha_s^2).
\eeq
Calculations of the one-loop corrections \cite{fie:81,dit:81} to the pion
form factor reveal that the magnitude of the $K$-factor strongly depends
on the renormalization scale. It is in general large except the renormalization
scale is chosen like $\mu = \sqrt{x_1 y_1} Q$ (see Sect.~1). For this choice
and the use of the asymptotic DA, K is about 1.3 in the few GeV region.
For DAs broader than the asymptotic one, i.e.~for such with a stronger
weight of the end-point regions, B seems to be negative. Note that at
least part of the K-factor is included in the Sudakov factor. With regard
to the new developments discussed above it is perhaps advisable to reanalyse
the one-loop corrections.\\
Disregarded soft contributions offer another explanation of the eventual
discrepancy between theory and experiment. As the $k_\perp$-effects
discussed above such contributions are of higher twist type and do not
respect the quark counting rules. Dominance of such contributions in the
case of the pion's form factor and perhaps in other exclusive quantities
would leave unexplained the apparent success of the counting rules.\\
There are several possible sources for such soft contributions:\\
i) Genuine soft contributions like VMD contributions or contributions
from the overlap of the soft parts of the hadronic wave functions.
The overlap contribution can be
estimated with the aid of the famous Drell-Yan formula \cite{Dre:70} (note
that the HSP represents the contribution from the overlap of the perturbative
tails of the hadronic wave functions). In the pion case it reads
\beq
\label{eq:Fpi-soft}
{F_\pi}^{soft}(Q^2)=\int \frac{dx_1 d^2k_{\perp}}{16 \pi^3}
   \,\Psi^{*}(x_1,\vec{k}_{\perp} + x_2\vec{q})\,\Psi(x_1,\vec{k}_{\perp})
\eeq
($Q^2=\vec{q}^2$). The integral is dominated by the region near $x_1=1$,
other regions are strongly damped by the wave function.
Hence $F_\pi^{soft}$ sensitively reacts to the behaviour of the wave function
for $x_1 \to 1$. Evaluation of (\ref{eq:Fpi-soft}) reveal that the MAS wave
function provides a soft contribution of the right magnitude to fill
in the gap between the perturbative contribution (\ref{eq:ft-Fpi}) and the
experimental data. The sum of the soft and the perturbative contribution
incidentally simulates the dimensional counting rule
behaviour in the region $1.5 - 3 \,{\rm GeV}$. Using instead of the Gaussian,
a $k_{\perp}$ dependence in accord with the moments
$\langle k_{\perp}^n\rangle $ derived from QCD sum rules the behaviour
$Q^2 F_\pi \simeq {\rm const}$ is found to hold even up to $6\,{\rm GeV}$
\cite{zhi:94}. Soft contributions of the type (\ref{eq:Fpi-soft}) are also
discussed in \cite{Isg:89,kis:94}.
Strong soft contributions to form factors are also obtained from
QCD sum rules \cite{rad:91,bra:94}.\\
ii) There may be orbital angular momentum components in the hadronic
wave function other than zero.
New phenomenological functions appear in general which is certainly a
disadvantage but may lead to a better quantitative description of the form
factor data. $L \neq 0$ components have the appealing consequence of violating
the helicity sum rule for finite values of $Q$ \cite{bol:93}. This may offer
a possibility to calculate the Pauli form factor of the proton.\\
iii) Contributions from higher Fock states are another source of higher twist
contributions.
Also in this case new phenomenological functions have to be introduced.\\
iv) For baryons one may also think of quark-quark correlations in the wave
functions which constitute higher twist effects. In a series of papers
(see \cite{pil:93,jkss:93} and references therein) the idea has been put
forward that such correlations can effectively be described by
quasi-elementary diquarks.
A systematic study of all exclusive photon-proton reactions has been
carried out in the diquark model, which is a variant of the unmodified HSP:
form factors in the space-like and in the time-like regions, virtual and
real Compton scattering, two-photon-annihilations into proton-antiproton
as well as photoproduction of mesons.
A fair description of all the data has been achieved utilizing in all
cases the same proton DA (as well as the same values for the
parameters specifying the diquarks). The diquark model
allows to calculate helicity flip amplitudes and consequently to predict
for instance the Pauli form factor of the proton. The results for it, shown in
Fig.~1, are in agreement with the data. Results for the magnetic form
factor in both the time-like and the space-like regions are shown in Fig.~5.
\bfig[p]
\vspace*{-0.4cm}
\caption[figdummy]{\small
The magnetic form factor of the proton in the time-like and space-like
(at $Q^2=-s$) regions. The time-like data are taken from
{\rm\cite{arm:93}}, the space-like data from {\rm\cite{sil:93}}.
The solid lines represent the predictions of the diquark model
{\rm\cite{pil:93}}, the dashed line is Hyer's prediction {\rm\cite{hy:93}}.}
\efig
\vspace*{-0.7cm}
\section{Summary}
\vspace*{-0.5cm}
The modified HSP which includes both the Sudakov corrections and the intrinsic
$k_\perp$-dependence of the hadronic wave function constitutes an
enormous progress in our understanding of exclusive reactions although
there are still some theoretical problems left. It provides an explicit
scheme for the infrared protection of the ``bare'' coupling constant.
The modified HSP
allows to calculate the perturbative contribution to form factors
in a theoretically self-consistent way for momentum transfers as low
as few GeV (about $2 (3)\,{\rm GeV}$ in the pion (nucleon) case). This is,
however, achieved at the expense of
strong suppressions of the perturbative contributions as compared to those
obtained with the original unmodified HSP. Now the perturbative contributions
are too small as compared to data. It thus seems that other contributions
(higher order corrections to the hard scattering amplitudes and/or higher
twists) also play an important role in the few GeV region as already
indicated by the recently measured Pauli form factor of the proton.
An interesting task for the future is to find out the size of such
higher twist contributions and to elucidate their physical nature.
Finally, I would like to emphasize that the approximate validity of the quark
counting rules, for which the HSP offers an explanation, would remain
a mystery if all large momentum transfer data for exclusive processes are
dominated by soft contributions.

\end{document}